# Potential of photon-subtracted CV states towards gain sensitivity of the Mach-Zehnder interferometer

## Mikhail S. Podoshvedov[1,2,3] and Sergey A. Podoshvedov[1,2]

[1]*Laboratory of quantum information processing and quantum computing, laboratory of quantum engineering of light, South Ural State University (SUSU), Lenin Av. 76, Chelyabinsk, Russia*
[2]*Laboratory of quantum engineering of light, South Ural State University (SUSU), Lenin Av. 76, Chelyabinsk, Russia*
[2]*Kazan Quantum Center, Kazan National Research Technical University named after A.N. Tupolev, Kazan, Russia*

**Abstract:** Quantum Cramer-Rao (QCR) bound is attached to a particular nonclassical state, therefore appropriate choice of the probe state is of the key importance to enhance sensitivity beyond classical one. Since the work of C.M. Caves (*Phys. Rev. D **23** 1693 (1981)*) Mach-Zehnder (MZ) interferometry operates with single-mode squeezed vacuum (SMSV) light coupled with a coherent state. We report the gain sensitivity of the phase-dependent MZ interferometer by more than $10\ dB$ compared to the original result (*Phys. Rev. Lett. **100**, 073601 (2008)*) by using SMSV state with squeezing $< 10\ dB$ from which a certain number of photons was initially subtracted. The gain sensitivity is also observed when measuring the difference of output intensities of the SMSV state with squeezing $< 3\ dB$ from which 2,4,6 photons are subtracted and large coherent state. Overall, subtracting photons from the initially weakly squeezed light can prove to be a more efficient strategy in the quantum MZ interferometry compared to highly squeezed SMSV state generation.

## 1. Introduction.

Measuring a physical system allows for one to extract part of information about its evolution over time [1]. A set of repeated measurements can lead to scattered output data because different physical mechanisms contribute to the measurement. Best measurement precision of probe classical system defined as the standard deviation (SD) of the estimated unknown parameter is limited by shot-noise scaling, i.e. by quantity which scales with reciprocal of square-root of the number of particles [2]. Probe nonclassical states can have additional interior resources to decrease the phase uncertainty with respect to classical one, beating the shot-noise limit (SNL) [3-6].

Sensitivity of the optical MZ interferometry with coherent states is limited to the standard quantum limit (SQL) or the same SNL. It can be assumed that increasing the intensity of the light fields in the MZ interferometer and, therefore, the mean number of photons, can significantly reduce its phase uncertainty. In a real situation, the radiation pressure fluctuations can affect the interferometric mirrors, making an additional contribution to the phase uncertainty thereby reducing the possible gain in sensitivity from increasing light intensity. The solution to the problem of estimate of unknown phase shift was proposed in the work by Caves [7] in which SMSV state [8] as nonclassical probe together with coherent one are directed to the MZ interferometer input. The work gave impetus to the development of phase-dependent optical quantum metrology [9-12]. The result [7] has been improved in the work [13] in which the authors considered a Bayesian phase inference protocol and have shown the phase sensitivity the MZ interferometer can reach the Heisenberg limit (HL) independently on the true value of the phase shift. Interferometric scheme with one beam splitter (BS) to evaluate the phase sensitivity with product state has been studied in [14]. It is known that if a vacuum is into input port of the MZ interferometer, then no matter what the state is in the other port and no matter what measurement is used, the sensitivity can never be better than the SQL [5,6]. Nevertheless, the phase uncertainty below SNL is observed with input port populated with Fock state while the second port is filled with the classic state [15]. Parity detection is proposed to realize sub-Heisenberg sensitivity in the MZ interferometer with two-mode squeezed vacuum state [16]. The prospects for performing a supersensitive estimate of the unknown phase difference using the MZ interferometer with input coherent and Fock states are presented in [17].

An interesting question remains whether it is still possible to improve sensitivity of the MZ interferometer beyond the limit provided by the SMSV state using quantum engineering of the probe state. The technique of subtracting and adding photons to the Gaussian states transforming them into non-Gaussian is well known [18-20]. Photon subtracted continuous variable (CV) states differ from the initial Gaussian state which can be useful for more precise parameter estimation [21,22]. Typically, $n-$photon subtraction is implemented by passing the corresponding state $\rho$ through a highly transmitting beam splitter, which allows the output $n-$ photon subtracted state to be approximated as $\rho_{-n} \sim a^n \rho a^{+n}$ in the case of recording the measurement outcome of $n$ photons in the neighboring measurement mode of the BS. From a theoretical point of view, such model can be conditionally realized with help of the BS with unit transmissivity. In relation to quantum optical metrology pure photon subtraction approach leads to rather modest gain in the interferometer sensitivity with respect to the initial CV state from which the photons are subtracted [23,24]. Moreover, the annihilation and creation operators are non-unitary and cannot be directly implemented in the laboratory experiment. Therefore, the practical approach based on exact solution of the problem of the passage of the SMSV state through an arbitrary BS with the hybrid entangled state at the output [25], deserves more attention since it provides the opportunity to take into account the redistribution of photons in various regimes of the BS, including also highly transmitting ones. It entails appearance of nonclassical CV states different from those that can only exist if the annihilation operators act on the initial state [26,27]. Gaussian states from which photons have been subtracted can exhibit new nonclassical properties [28], in particular, photon subtraction can increase the brightness of the measurement-induced CV state [29] and also generate squeezing-enhanced states [30] i.e., those which have more squeezing than the original SMSV state. Another aspect of the practical theory of the multiphoton distribution on a beam splitter with variable transmissivity can be related to the deterministic implementation of the hybrid entanglement [31,32] as well as quantum teleportation of the photonic state with the possibility of restoring the transmitted qubit with a pre-known amplitude distorting factor [33]. Experimental demonstration of quantum state engineering in which quantum fluctuations, mean photon numbers and correlations are manipulated is demonstrated in [34,35]. A modified version may involve subtracting the displaced photon state by preliminary using the displacement operator [36]. Progress in the development of photon-number resolving detection based on use of some highly efficient transition edge sensors (TESs) [37] makes it possible to apply the technology to the subtraction of large number of photons. As for the generation of the squeezed light, the realized squeezing ranges from $9 \ dB$ in [38] and $10 \ dB$ in [39] up to $15 \ dB$ at some near unfrared wavelengths [40]. Some aspects of the squeezed states with a focus on experimental observations are presented in [41].

Here we consider the metrological potential of the photon-subtracted CV states of a certain parity together with a coherent state in the quantum MZ interferometry. Weakly squeezed states are not very widespread, in particular, due to some their closeness to coherent states [41]. Losses of photons may bring a squeezed light closer to a coherent one. Subtracting photons from the weakly squeezed light to implement a probe state for MZ interferometry turns out to be more effective approach compared to the generation of highly squeezed light. The measurement-induced CV states of a certain parity can carry larger quantum Fisher information (QFI) compared to the original probe state which increases their metrological capabilities. Subtraction and addition of photons from the SMSV state in order to increase the sensitivity of the MZ interferometer is considered in [42], limited only to cases of maximum addition and subtraction of 3 photons. Naturally there is no reason not to consider a larger number of subtracted photons (say more than 3) which could lead to even greater gain in the sensitivity of the MZ interferometer. A realistic model with BS parameters that can be changed allows one to estimate the ultimate phase uncertainty for a sufficiently large photon subtraction which may become practical for initially weakly squeezed states. Given the decrease in the limiting boundary of the phase uncertainty with increasing number of subtracted photons, we also consider the phase estimate with less uncertainty when measuring the intensity difference at the output of the MZ interferometer which also gives better results compared to using the original SMSV state.

## 2. Gain of sensitivity via photon subtraction.

The original interferometric scheme in context of gravitational wave detection [7] is sketched in Fig. 1. It is pumped with a probe state

$$|\Psi_{SMSV,\alpha}(y,\alpha)\rangle_{12} = |SMSV(y)\rangle_1 |i\alpha\rangle_2, \tag{1}$$

being product of the SMSV state in mode 1 written here as

$$|SMSV(y)\rangle = \frac{1}{\sqrt{\cosh s}} \sum_{n=0}^{\infty} \frac{y^n}{\sqrt{(2n)!}} \frac{(2n)!}{n!} |2n\rangle, \tag{2}$$

where an amount $y = \tanh s/2$ is a squeezing parameter, $s > 0$ is the squeezing amplitude which provides the range of its change $0 \le y \le 0.5$ and coherent state $|i\alpha\rangle$ in mode 2 with real amplitude $\alpha > 0$. The SMSV state is produced by parametric down conversion [41]. This is a second order nonlinear effect in which pump photons are converted to photons of less frequency creating a superposition of even number states (2). In addition, the SMSV state can be characterized by the squeezing parameter $S = -10 \, log\big(exp(-2s)\big) \, (dB)$ and the average number of photons $\langle n_{SMSV}\rangle = sinh^2 s$.

Maximal phase sensitivity $\Delta\varphi$ of the MZ interferometer illuminated by the probe state (1) is given by QCR boundary [13]

$$\Delta\varphi_{SMSV,\alpha} = \frac{1}{\sqrt{F_{SMSV,\alpha}(s,\alpha)}} = \frac{1}{\sqrt{\alpha^2 exp(2s) + sinh^2(s)}}, \tag{3}$$

where the QFI being just four times variance of phase evolution generator $J_y$, can also be represented in a compact form $F_{SMSV,\alpha}(s,\alpha) = 2(1 + \operatorname{ctanh} s)\langle n_{SMSV}\rangle\alpha^2 + \alpha^2 + \langle n_{SMSV}\rangle$ in terms of the average number of particles. The QCR bound is independent on the true value of the phase shift and can show sub-SQL $\Delta\varphi \sim 1/\sqrt{n}$, where $n$ is an average number of input particles.

Here we are interested in finding a probe state that could show greater sensitivity compared to that presented in equation (3)

$$|\Psi_{n,\alpha}(y_1,\alpha)\rangle_{12} = |\Psi_n^{(0)}(y_1)\rangle_1 |i\alpha\rangle_2, \tag{4}$$

in which the SMSV state is replaced by the one from which $n$ photons have been subtracted. The CV states of definite parity can be realized through a measurement-induced mechanism implemented on the basis of the beam splitter and a PNR detector [25]. The form of measurement-induced states as well as the notations $y_1, n, (0)$ used are presented in [26,27].

Minimal uncertainty of the phase estimate with probe state in equation (4)

$$\Delta\varphi_{n,\alpha} = \frac{1}{\sqrt{F_{n,\alpha}(s,\alpha)}}, \tag{5}$$

follows from their QFI

$$F_{n,\alpha}(s,\alpha) = 2(1 + 2y_1)\langle n\rangle\alpha^2 + \big(1 + 4y_1(n+1)\big)\alpha^2 + \langle n\rangle. \tag{6}$$

Here there $\langle n\rangle$ is the mean number of photons in $n$−photon subtracted CV state compact form of which is present in [29] , while $\langle n_\alpha\rangle = \alpha^2$ is an average number of photons in the coherent state.

In Figures 2(a-d) we show the dependence of the QCR bound on the squeezing $S \, (dB)$ of the original SMSV state both for the probe state (1) (highlighted in blue) and state (4) with some $n$ for different values of coherent amplitude $\alpha = 1$ (Fig. 2(a,b)) and $\alpha = 100$ (Fig. 2(c,d)). Graphs in Figure 2 (a,c) are plotted for the BS parameter $t = 0.9$ and for $t = 0.99$ in Figure 2(b,d). Overall, the following features can be marked in the graphical dependencies. Firstly, an increase of the squeezing parameter $S$ leads to a monotonous increase of the limiting sensitivity no matter which CV states of certain parity are used as probe. Secondly, at the initial stage of $S$ which starts from $S_1 = 0$, the maximal sensitivity of the probe state in Eqs. (4), is ahead and only at a certain sufficiently large value of $S = S_2$ the QCR bounders of the probe states cross with further advance of the state in formula (1). It clearly indicates in favor of a preliminary procedure of subtracting photons from feebly squeezed SMSV state. The value of the parameter $S_2$ depends in a rather complex way both on the amplitude $\alpha$ of the coherent state, number $n$ of subtracted photons, input squeezing amplitude $S$ and the transmittance $t$ of the BS with the help of which $n -$ photon subtracted CV state is realized. Regardless of the values of the parameters, the values of $S_2$ are in the region of practically realizable SMSV states (say, in general, $S_2 > 15\ dB$). Third, an increase in number of photons subtracted guarantees the movement of the parameter $S_2$ towards larger values along the horizontal axis. Fourthly, increasing the amplitude of coherent state from $\alpha = 1$ up to $\alpha = 100$ also allows for one to significantly increase the phase sensitivity of the MZ interferometer. The increase in sensitivity occurs quite monotonously with increasing $\alpha$. Fifth, the QCR boundary of the probe state (4) can also depend on BS parameter $t$. The functions comparable to the HL as well as SQL, i.e., $HL = 1/(\langle n \rangle + \alpha^2)$ and $SQL = 1/\sqrt{\langle n \rangle + \alpha^2}$ for $n = 80$ are also presented in Figs. 2. Basically, the QCR boundaries lie in the range between SQL and HL, i.e. $SQL > \Delta\varphi_{n,\alpha} > HL$, which indicates overcoming SQL. But at small values of $\alpha$ (say $\alpha = 1$), QCR bounds can show sub-Heisenberg scaling, i.e. $HL > \Delta\varphi_{n,\alpha}$ the case not shown in Figs. 2.

To show quantitatively a gain sensitivity of the probe state (4) with respect to the original state in Eq. (1), we present parameter $g_n = -10\ lg\left(\Delta\varphi_{n,\alpha}/\Delta\varphi_{SMSV,\alpha}\right)(dB)$ in Figs. 3 in dependence on the input squeezing amplitude $S\ (dB)$. The gain sensitivity experiences maximum at some value $S$. The highs and their shift along the horizontal axis $S$ depends on $n$, $\alpha$ and $t$. What is interesting is that the gain sensitivity depends to a much greater extent on $t$ than on $\alpha$. In general, it is possible to select the values $n$, $\alpha$ and $t$ that could provide the gain sensitivity $> 10\ dB$ just for practically feasible small values of the squeezing amplitude $S < 10\ dB$, in particular, at $t = 0.9$, the maximum gain sensitivity is observed at $S \approx 5\ dB$.

The potential that $n -$photon subtracted CV states can demonstrate when estimating an unknown phase shift is worth testing by measuring the intensity difference at the output of the MZ interferometer following error propagation formula [4] for the phase uncertainty

$$\Delta\varphi = \frac{\Delta J_z}{\left|\frac{\partial \langle J_z \rangle}{\partial \varphi}\right|} = \frac{\Delta J_z}{sin\varphi|\langle J_z \rangle|}. \tag{7}$$

Optimal point for estimate follows from phase shift $\varphi = \pi/2$. Numerical uncertainties $\Delta\varphi$ in dependency on squeezing amplitude $S$ are shown in Figure 4(a,b) for $\varphi = \pi/2$ and different values of $n$, $\alpha$ and $t$ both for the probe states (1) [7] and (4). In the case of a small amplitude $\alpha = 1$ of the coherent state, the sensitivity of the MZ interferometer with probe states (4) clearly surpass the one of the original state (1) with squeezing $S < 20\ dB$. If the amplitude of the coherent state increases, the situation becomes less clear. The range of superior sensitivity of the state (4) is reduced but nevertheless exists even for $\alpha = 100$. What is important is that superiority is observed for squeezed SMSV states with squeezing $< 3\ dB$ and for small values of subtracted photons $n = 2,4,6$. The expression for the uncertainty of the phase shift (7) contains in the denominator the difference of the mean number of photons. Subtracting photons redistributes the output distribution towards increasing the average number of photons, which brings it closer to the average number of photons in the coherent state which guarantees growth in uncertainty $\Delta\varphi$ with $S$ increasing.

## 3. Conclusion.

In conclusion, we have shown that the metrological potential of the MZ interferometry can increase with probe state in Eq. (4) even achieving sub-Heisenberg sensitivity. The gain in sensitivity compared to the original probe state in equation (1) with the squeezing $< 10\ dB$ can exceed $10\ dB$ which raises the practical significance of the photon subtraction approach. Thus, the generation of a highly squeezed state can be replaced by a more efficient procedure of photon subtraction from weakly squeezed SMSV state. The gain sensitivity is also attained when measuring the output intensity difference, even though the measurement does not provide reaching the QCR bound. The gain sensitivity in the case of $\alpha = 100$ is realized in the case of using SMSV state with squeezing $< 3\ dB$ from which $n = 2,4,6$ photons is subtracted. So, we have $g_4 = 1.35\ dB$ at initial squeezing $S = 1.53\ dB$ and $g_6 = 1.49\ dB$ with $S = 1.02\ dB$. Given the small number of photons required to subtract, the approach is feasible and can be tested in practice.

**Acknowledgement**

The work of MSP and SAP was supported by the Foundation for the Advancement of Theoretical Physics and Mathematics "BASIS".

**Disclosures**

The authors declare no conflicts of interest.

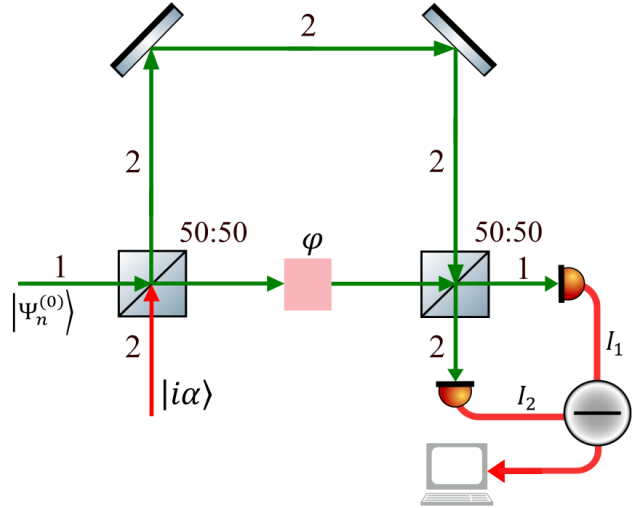

Fig. 1. MZ interferometer, on one input of which (first) photon-subtracted CV state is launched while the coherent state $|i\alpha\rangle$ ($\alpha > 0$ is real) is applied to the second port. Phase shift $\varphi = \pi/2$ can be fairly accurately estimated by measuring the difference of the output intensities. Subtracting photons from weakly squeezed states is a more efficient approach compared to generating highly squeezed SMSV state.

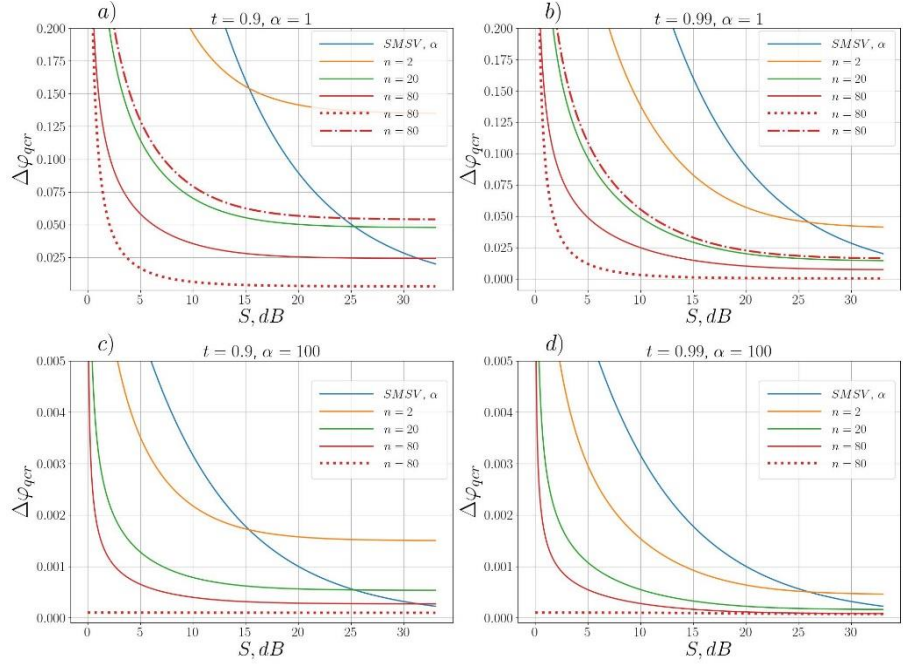

Fig. 2(a-d). Dependence of the limiting bound $\Delta\varphi_{qcr}$ (here and in other figures abbreviation $qcr$ applies in subscript) on the squeezing $S$ $(dB)$ of the original SMSV state for different number of subtracted photons $n$, coherent state amplitudes $\alpha$ and transmittance $t$ of the beam splitter used in quantum engineering of the CV states of definite parity. In addition, the SQL curve $SQL = 1/\sqrt{\langle n \rangle + \alpha^2}$ (dash-dot lines) and the HL curve $HL = 1/(\langle n \rangle + \alpha^2)$ (dashed lines) for $n = 80$ are also shown. SQL curves for $\alpha = 100$ are not shown since they lie significantly above.

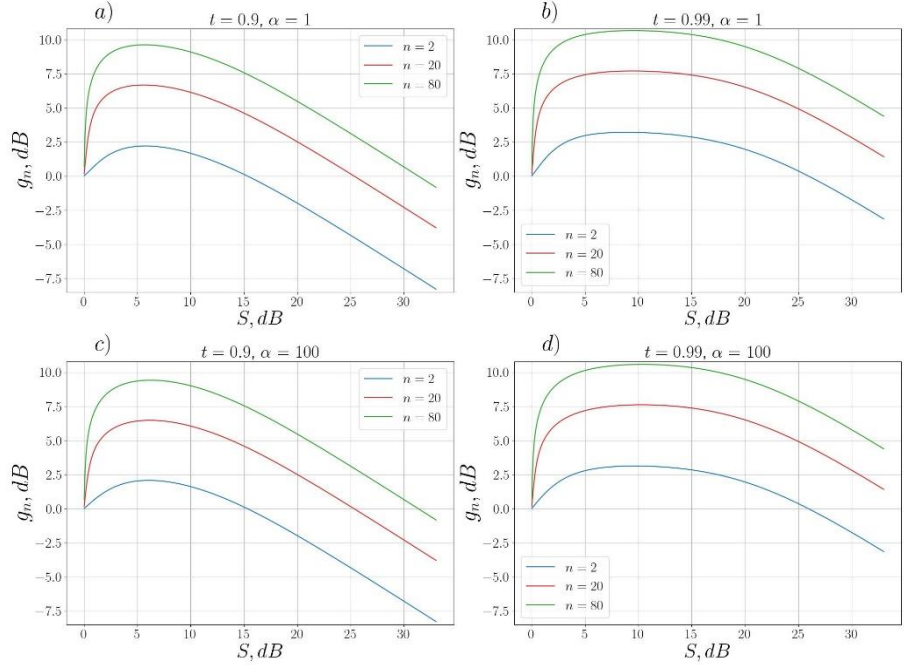

Fig. 3(a-d). Gain sensitivity of the MZ interferometer with a probe state (4) in dependency on the squeezing amplitude $S$ $(dB)$ of the original SMSV state for different values of $n$, $\alpha$ and $t$.

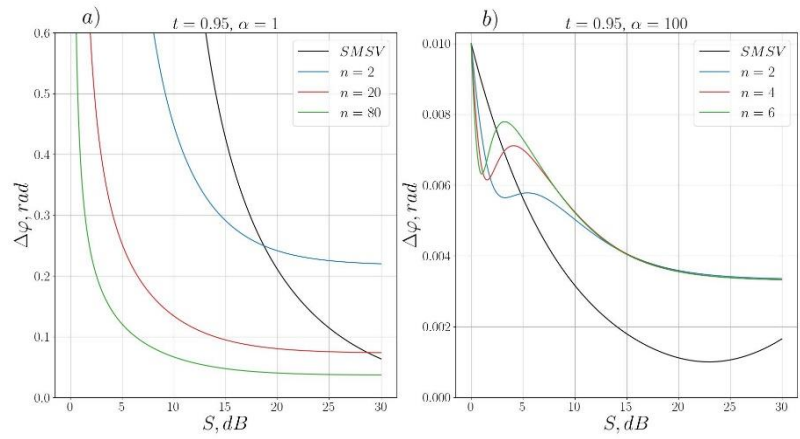

Fig. 4(a,b). Uncertainty $\Delta\varphi$ of the phase shift $\varphi = \pi/2$ estimated by the difference in intensities at the output of the MS interferometer with probe states (1) and (4) as function of the squeezing of original SMSV state. Graphs are plotted for various values of $\alpha$ and $t$. Even for large amplitude $\alpha = 100$ of the coherent state there is an increase in sensitivity in the region of $< 3\ dB$ of input squeezing in the case of a small number ($n = 2,3,4$) of subtracted photons.